\documentclass[a4paper, oneside, twocolumn, notitlepage, 10pt]{extarticle_ecoc}
\usepackage{ecoc}

\begin{document}
\selectlanguage{english}    


\title{Fully Distributed Fiber-Optic Sensing Enabled by Kalman Filtering}%


\author{
    Juan M. Marin\textsuperscript{(1)}, Roman Ermakov\textsuperscript{(1)},
    Florian Azendorf\textsuperscript{(2)}, André Sandmann\textsuperscript{(2)}, \\ Francesco Da Ros\textsuperscript{(1)}, Darko Zibar\textsuperscript{(1)}
}

\maketitle                  


\begin{strip}
    \begin{author_descr}

        \textsuperscript{(1)} Department of Electrical and Photonics Engineering, Technical University of Denmark, Kongens Lyngby 2800, Denmark, \textcolor{blue}{\uline{jmmmo@dtu.dk}}

        \textsuperscript{(2)} Adtran Networks SE, Märzenquelle 1-3, 98617 Meiningen, Germany

    \end{author_descr}
\end{strip}

\renewcommand\footnotemark{}
\renewcommand\footnoterule{}


\begin{strip}
    \begin{ecoc_abstract}
Signal fading creates points along the fiber where phase cannot be extracted, so they are conventionally discarded. Instead, we propose a Kalman-based solution for $\phi$-OTDR full-fiber monitoring. Experiments demonstrate phase and temperature estimation with approximately 15 times better spatial density than if removing these points.
    \end{ecoc_abstract}
\end{strip}


\section{Introduction}

\noindent Phase-sensitive optical time-domain reflectometry ($\phi$-OTDR) monitors tens of kilometers of fiber by detecting phase variations in Rayleigh backscattering~\cite{Healey1984}. When monitored over consecutive time periods, the resulting interference patterns encode external perturbations, revealing a spatiotemporal map of the fiber's environment~\cite{muanenda2018, Liokumovich2015}. The relatively high intensity of Rayleigh backscattering enables longer sensing range and faster measurements than other OTDR-based technologies~\cite{lu2019}. Additionally, $\phi$-OTDR can be integrated into deployed telecom fibers, enabling reliable and context-aware optical networks by providing them with sensing capabilities~\cite{kaszubowska2025, ip2022}. Consequently, substantial research has focused on $\phi$-OTDR for strain~\cite{hartog2015} and, more recently, temperature sensing~\cite{ermakov2025, wang2025}.

Nevertheless, $\phi$-OTDR still faces challenges, such as polarization fading. This phenomenon occurs when the backscattering polarization becomes orthogonal to that of the local oscillator~\cite{vidal2023}. Therefore, the resulting beat signal suffers attenuation at arbitrary points, resulting in low signal-to-noise ratio (SNR)~\cite{gabai2016}. Conventionally, these points are removed~\cite{ermakov2025}, creating blind spots along the fiber's length. In turn, this can pose serious risks in applications that monitor critical infrastructure, such as telecom networks~\cite{sadighi2025}, power cables~\cite{yilmaz2006} and oil-and-gas pipelines~\cite{ashry2022}. In these contexts, early and accurate hotspot detection is crucial to preventing failures leading to catastrophic consequences. To address this issue, polarization diversity and multi-frequency detection schemes have been explored~\cite{martins2016, zhou2013}. However, these approaches require additional components, increasing costs and hardware complexity.

In this work, we propose a solution that recursively estimates phase by combining noisy measurements with physics-informed predictions. To achieve this, we designed a framework based on the Sage–Husa adaptive Kalman filter (SHAKF)~\cite{sage1969} capable of tracking the phase distribution at each point along the time (slow) and spatial (fast) axes. This enables the prediction of phase at faded points by following the distribution of all preceding points. We experimentally verify our approach by estimating phase and temperature from $\phi$-OTDR measured data, demonstrating that all measured points along the fiber can be used for sensing.

\section{Materials and Methods}

\begin{figure*}[t]
    \centering
    \includegraphics[width=\textwidth]{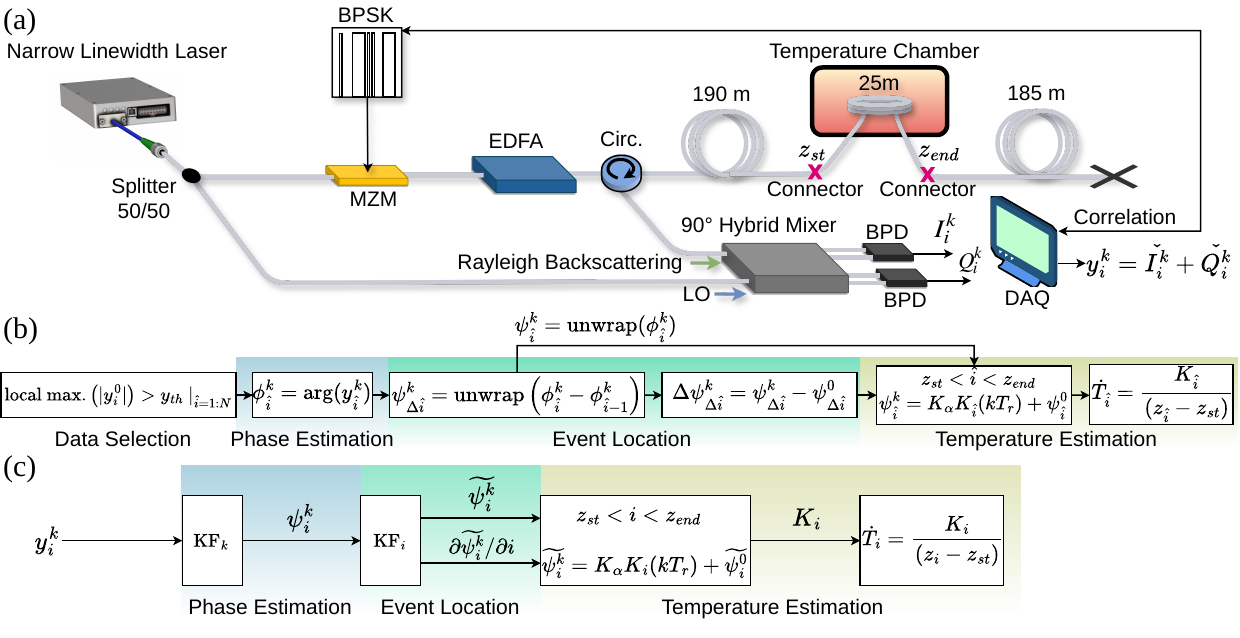}
    \caption{(a) Experimental setup for $\phi$-OTDR. Schematics of (b) conventional and (c) Kalman-based temperature estimation.}
    \label{fig:1}
\end{figure*}

Figure~\ref{fig:1}(a) illustrates the $\phi$-OTDR experimental setup. The signal power of a narrow-linewidth laser is split by a 50/50 coupler into a probe signal path and a local oscillator (LO) path. In the signal path, a Mach–Zehnder modulator (MZM), driven at 125~Mbit/s, encodes the light into binary phase-shift keying (BPSK) pulses using a 4095-bit pseudo-random sequence with 5000-symbol zero-padding. This yields a frame duration $T_{r}$ of 72.8~$\mu$s and a spatial resolution of 80~cm. The modulated signal is amplified by an erbium-doped fiber amplifier (EDFA) and launched into the fiber under test (FUT) via an optical circulator. The FUT consists of two fiber sections (190~m and 185~m) connected by a 25~m patch cord placed inside a temperature-controlled chamber. The backscattered signal is combined with the LO at a 90$^\circ$ hybrid mixer, where each output is detected by a balanced photodetector (BPD) to recover in-phase and quadrature (I/Q) components. These are sampled at 625~MS/s using a digital acquisition system (DAQ). A correlation step then reconstructs the response of a single optical pulse from the transmitted sequence. A total of 646 time frames were recorded over 50~ms.

Using this data, we compare conventional phase and temperature estimation after removing faded points, as summarized in Fig.~\ref{fig:1}(b), with our proposed approach described in Fig.~\ref{fig:1}(c). Both methods follow three steps: phase estimation, event location, and temperature estimation. Each of these steps is explained and compared in detail on the next sections, and the achieved performance is evaluated at each of them.

\begin{figure}[t]
    \centering
    \includegraphics[width=\linewidth]{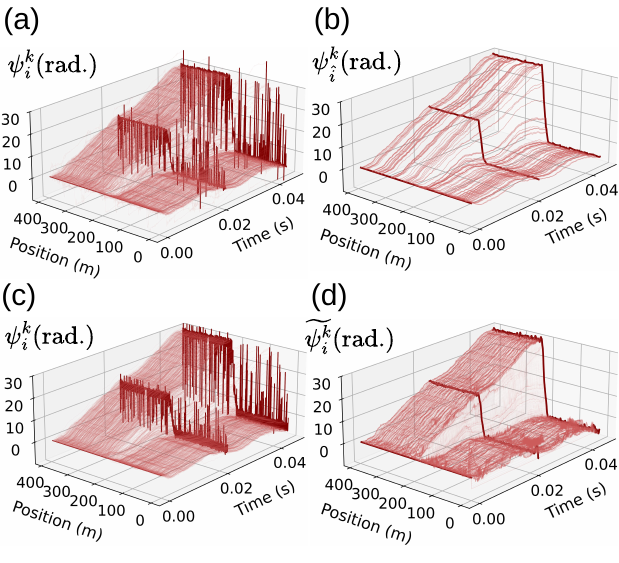}
    \caption{Phase visualization: (a) directly from I/Q argument, (b) after conventional faded point removal. Compared to (c) $\text{KF}_k$ phase estimations, and (d) after $\text{KF}_i$ spatial refinement.}
    \label{fig:2}
\end{figure}

\section{Phase Estimation}

The output of $\phi$-OTDR consists of I/Q components: $y_i^k = \check{I_i^k} + j\check{Q_i^k}$, where $(i,k)$ index position along the FUT and time, respectively. To illustrate the effect of fading points, we compute $\psi_i^k = \text{unwrap}\left(\text{arg}(y_i^k)\right)$ for every position along the FUT. The results are displayed in Fig.~\ref{fig:2}(a). Discontinuities along the $i$-axis indicate locations affected by polarization fading, which obscure the spatial phase distribution.

Conventionally, as described in Fig.~\ref{fig:1}(b), these points are removed. To this end, the indices of local maxima $\hat{i}$ in the initial fingerprint $|y_i^0|$ whose intensity exceeds a threshold $y_{th}$ are selected. The threshold is set here to the average intensity of those maxima, $\overline{|y_{\hat{i}}^0|}$. The resulting spatiotemporal phase distribution $\psi_{\hat{i}}^k$ is visualized in Fig.~\ref{fig:2}(b), showcasing the suppression of discontinuities at the expense of a significant reduction in the density of phase estimations along the fiber. 

Alternatively, the proposed approach, as illustrated in Fig.~\ref{fig:1}(c), employs a SHAKF, denoted $\text{KF}_k$, whose state vector $x_k = (A_{i-1}^k, \psi_{i-1}^k, A_i^k, \psi_i^k)$ tracks the amplitude and phase at consecutive points along the fiber. The state predictions follow a random walk $x_k = x_{k-1} + w_k$, with process noise $w_k \sim \mathcal{N}(0, Q)$. The filter relates its state predictions at time frame $k$ to the measurements $\hat{z}_k = [y_{i-1}^k, y_i^k]$ via the measurement function

\begin{equation}
    h(x_k)= \left[ \begin{matrix}
        A_{i-1}^k \cos (\psi_{i-1}^k)+jA_{i-1}^k\sin (\psi_{i-1}^k)\\
        A_i^k \cos (\psi_i^k)+jA_i^k\sin (\psi_i^k)
    \end{matrix}\right] + r_k,
\end{equation}

\noindent where $r_k \sim \mathcal{N}(0, R_k)$ is the measurement noise. Its covariance matrix $R_k$ is estimated iteratively at each step $k$ as a combination of the previous estimate $R_{k-1}$ and the innovation $\epsilon_k = \hat{z}_k - h(x_k)$,

\begin{equation}
R_k = (1-d_k)R_{k-1} + d_k(\epsilon_k^T \epsilon_k), \quad d_k = \frac{1-b}{1-b^{k+1}}.
\end{equation}

The forgetting factor $b$ balances measurements and predictions. We set $b=0.05$, placing high trust in predictions at $k=0$ while shifting to 95\% trust in measurements over time, ensuring responsiveness to phase changes. Figure~\ref{fig:2}(c) shows the phase estimates $\psi_i^k$ from $\text{KF}_k$. Although $\text{KF}_k$ tracks temporal dynamics well, it fails to resolve spatial discontinuities. This motivates the spatial-domain SHAKF introduced in the next section, which refines predictions by leveraging correlations between adjacent points. The results shown in Fig.~\ref{fig:2}(d) validate its ability to mitigate the impact of signal fading in phase estimation.
\begin{figure}[t]
    \centering
    \includegraphics[width=\linewidth]{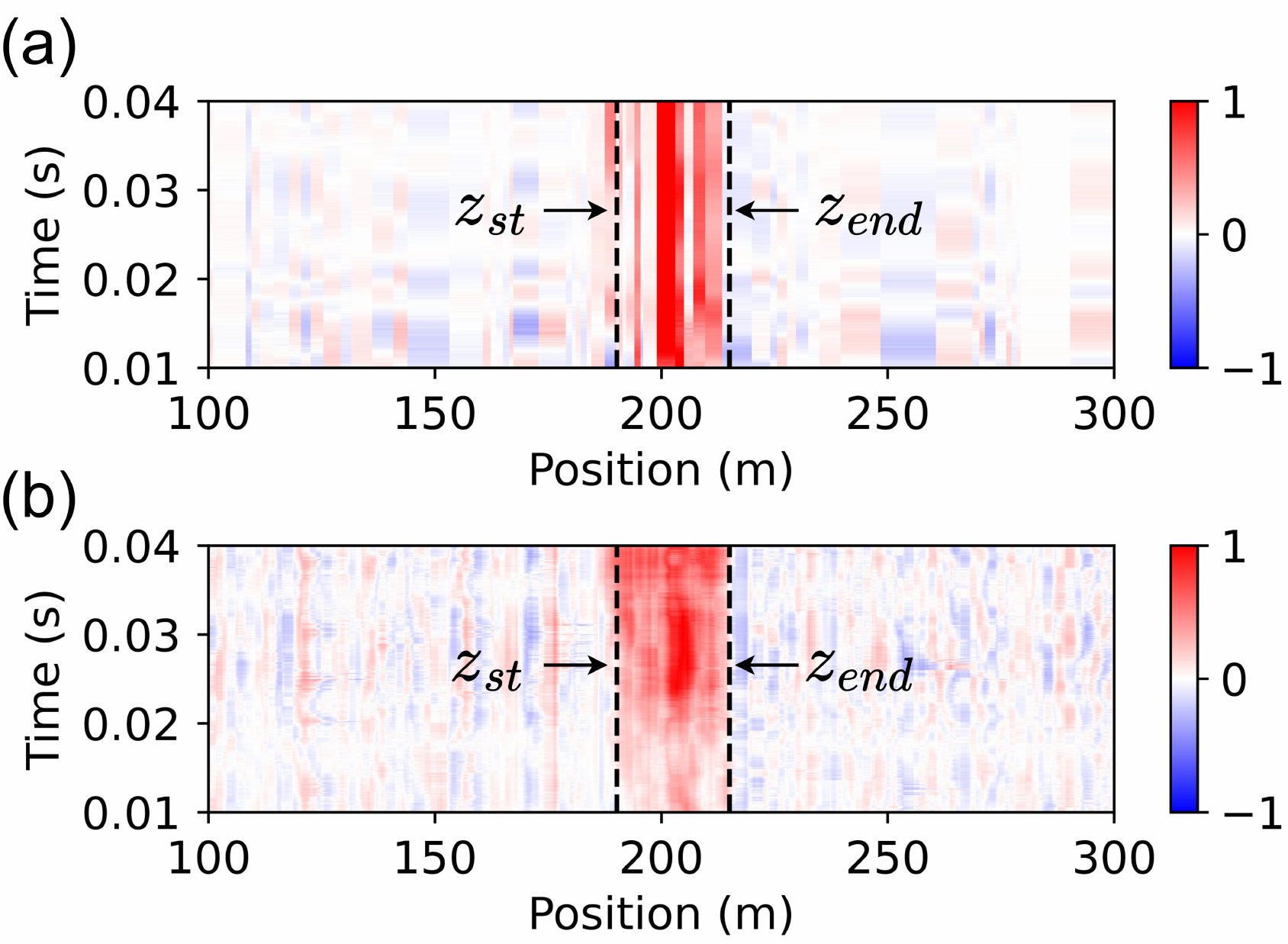}
    \caption{Normalized waterfall plots of (a) $\Delta \psi_{\Delta \hat{i}}^k$ and (b) $\partial \widetilde{\psi}_i^k /\partial i$.}
    \label{fig:3}
\end{figure}

\section{Event Location}

Before estimating temperature, the second stage of the pipeline locates the temperature event by identifying its start and end points $\left( z_{st}, z_{end}\right)$. Leveraging the temperature–phase relationship from~\cite{ermakov2025}, a heated fiber segment exhibits a linear phase evolution along the slow axis. This phase change is proportional to the cumulative temperature rate $K_i$ between $z_{st}$ and $z_{end}$. As a result, a spatial correlation 
emerges, manifesting as a phase slope between $z_{st}$ and $z_{end}$ that grows at each time step. Therefore, by analyzing the phase profile along the fiber ($i$-axis), we can clearly identify the region affected by temperature.

To achieve this, the conventional framework in Fig.~\ref{fig:1}(b) suggests monitoring the phase difference between consecutive selected points, denoted $\psi_{\Delta \hat{i}}^k$ and monitor the temporal evolution with respect to the initial frame: $\Delta \psi_{\Delta \hat{i}} ^k= \psi_{\Delta \hat{i}}^k - \psi_{\Delta \hat{i}}^0$. Fig.~\ref{fig:3}(a) shows the spatial distribution of $\Delta \psi_{\Delta \hat{i}}^k$ normalized at each time frame. In this figure, a hotspot is evident between the start and end of the 25~m fiber section placed inside the temperature chamber. 

Instead, to leverage all sampled points, our proposed approach in Fig.~\ref{fig:1}(c) introduces a second SHAKF, denoted as $\text{KF}_i$. It receives the phase estimates $\psi_{i}^k$ from $\text{KF}_k$ for every sampled position $z_i$ in the fiber, and models their spatial evolution in terms of the state vector $x_i = \left(\widetilde{\psi}_i^k, \partial \widetilde{\psi}_i^k/ \partial i\right)$ composed of the spatially smoothed phase and its spatial change of rate as 

\begin{equation}
   \left[ \begin{matrix}
    \widetilde{\psi}_{i+1}^k \\ 
    \partial \widetilde{\psi}_{i+1}^k/ \partial i
    \end{matrix}\right] = \left[ \begin{matrix}
    1 & 1 \\
    0 & 1
    \end{matrix}\right] \left[ \begin{matrix}
    \widetilde{\psi}_i^k \\ 
    \partial \widetilde{\psi}_i^k/ \partial i
    \end{matrix}\right]  + w_k.
\end{equation}

Additionally, $\text{KF}_i$ gives higher weight to predictions by setting $b=0.95$, enabling the filter to adjust to the phase spatial distribution, as evidenced in the resulting predictions of $\widetilde{\psi}_i^k$, exhibited in Fig.~\ref{fig:2}(d), and $\partial \widetilde{\psi}_i^k/ \partial i$ depicted in Fig.~\ref{fig:3}(b), demonstrating that it is not necessary to eliminate faded spots to accurately locate the event.


\section{Temperature Estimation}

\begin{figure}[t]
    \centering
    \includegraphics[width=\linewidth]{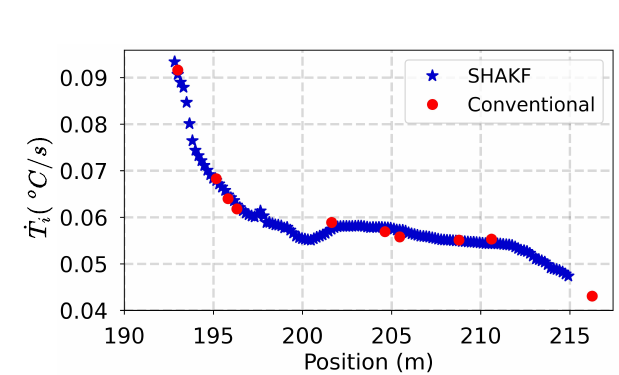}
    \caption{Estimated $\dot{T_i}$ along the heated fiber section.}
    \label{fig:4}
\end{figure}

A direct link between phase evolution and temperature dynamics was established in~\cite{ermakov2025}. In the temporal domain, the phase follows the linear approximation $\psi_i^k = K_{\alpha}K_i (kT_r) + \psi_i^0$, where $K_\alpha$ is the fiber's thermo-optical proportionality constant, $K_i$ is the cumulative temperature rate, and $\psi_i^0$ is the initial phase at $k=0$. Using linear regression, $K_i$ can be estimated for $z_{st} < z_i < z_{end}$. Assuming a rectangular temperature spatial profile, $K_i$ simplifies to

\begin{equation}
K_i = \int_{z_{st}}^{z_i} \dot{T}(x)\ dx \simeq \dot{T_i} (z_i - z_{st}),
\end{equation}

\noindent enabling direct estimation of the effective temperature change rate $\dot{T_i}$ in the time-domain for every $z_i$. Fig.~\ref{fig:4} compares the thermal distribution obtained with the conventional framework (Fig.~\ref{fig:1}(b)) versus the SHAKF framework (Fig.~\ref{fig:1}(c)). The results show a substantial improvement in sampling density without compromising measurement fidelity. The latter is evident by comparing the averaged $\dot{T_i}$ measurements with the chamber's thermostat reading, representing an average $\dot{T_i}$ of 0.06$^\circ$C/s inside the chamber. The conventional framework yields only 10 monitoring points along the 25 m heated fiber section, with an average $\dot{T_i}$ of 0.061 $^\circ$C/s. In contrast, our approach provides 146 points with an average of 0.063 $^\circ$C/s, showcasing a significantly denser and smoother temperature profile while still preserving an averaged temperature close to the thermostat reading.

\section{Conclusions}

We have presented a novel framework based on Kalman filtering that addresses signal fading in $\phi$-OTDR by analyzing spatiotemporal phase evolution. Our results demonstrate that this approach enables accurate phase extraction from all points along the fiber. Whereas, conventional phase estimation requires removing faded points, yielding only a fraction of the spatial density. This was evident as our approach achieved approximately 15 times better spatial density. These results highlight Kalman filtering as a pathway to reliable, cost-efficient, and fully distributed fiber sensing.

\clearpage
\section{Acknowledgements}
This work is funded by ICON (HORIZON Project 101189703) and Villum Fonden (VI-POPCOM VIL54486, OPTIC-AI VIL29334).

\defbibnote{myprenote}{%
Citations must be easy and quick to find. More precisely:
\begin{itemize}
    \item Please list all the authors. 
    \item The title must be given in full length. 
    \item Journal and conference names should not be abbreviated but rather given in full length.
    \item The DOI number should be added incl. a link.
\end{itemize}
}
\printbibliography[]

\vspace{-4mm}

\end{document}